\documentclass[fleqn,10pt]{wlscirep}
\usepackage[utf8]{inputenc}
\usepackage[T1]{fontenc}
\title{Optimal COVID-19 infection spread under low temperature, dry air, and low UV radiation}

\author[1]{Eitan E. Asher}
\author[2,*]{Yosef Ashkenazy}
\author[1]{Shlomo Havlin}
\author[3]{Alon Sela}
\affil[1]{Department of Physics, Bar-Ilan University, Ramat-Gan 52900, Israel}
\affil[2]{Department of Solar Energy and Environmental Physics, The Blaustein Institutes for Desert Research, Ben-Gurion University of the Negev, Midreshet Ben-Gurion, 84990, Israel}
\affil[3]{Industrial Engineering Department, Ariel University, Ariel, Israel}

\affil[*]{ashkena@bgu.ac.il}


\begin{abstract}
  The COVID-19 pandemic, caused by the novel coronavirus SARS-CoV-2, is spreading rapidly throughout the world, causing many deaths and severe economic damage. It is believed that hot and humid conditions do not favor the novel coronavirus, yet this is still under debate due to many uncertainties associated with the COVID-19 data. Here we propose surrogate data tests to examine the preference of this virus to spread under different climate conditions. We find, by mainly studying the relative number of COVID-19 deaths, that the disease is significantly (above the 95\% confidence level) more common when the temperature is $\sim$10$^\circ$C, the relative humidity is $\sim$60\%, the specific humidity is $\sim$5 g/kg, and the ultraviolet (UV) radiation is $\sim$80 kJ/m$^2$. The results are supported using global and regional data, spanning the time period from January to August 2020. The COVID-19 data includes the daily reported new cases and daily death cases; for both, the population size is either taken into account or ignored.
\end{abstract}
\begin{document}

\flushbottom
\maketitle
%
%
\thispagestyle{empty}


\section*{Introduction}

The COVID-19 pandemic has quickly spread throughout the entire world, with a high cost in lives and severe economic damage. As of September 2, 2020, the novel coronavirus has infected over 26 million people (confirmed cases) and caused 0.86 million deaths globally. Since the disease mainly spreads by human breath through air droplets, and with the current lack of a vaccine or effective treatments, the main practices to prevent the spread of COVID-19 include social distancing, sanitation, masks, and reducing human mobility by adopting quarantine policies. However, full mobility limitations in an entire country can only be made for short periods, mainly due to high economic costs\cite{fernandes2020economic, nicola2020socio}. 

\begin{figure}[htb!]
	\centering
	\includegraphics[width = \textwidth]{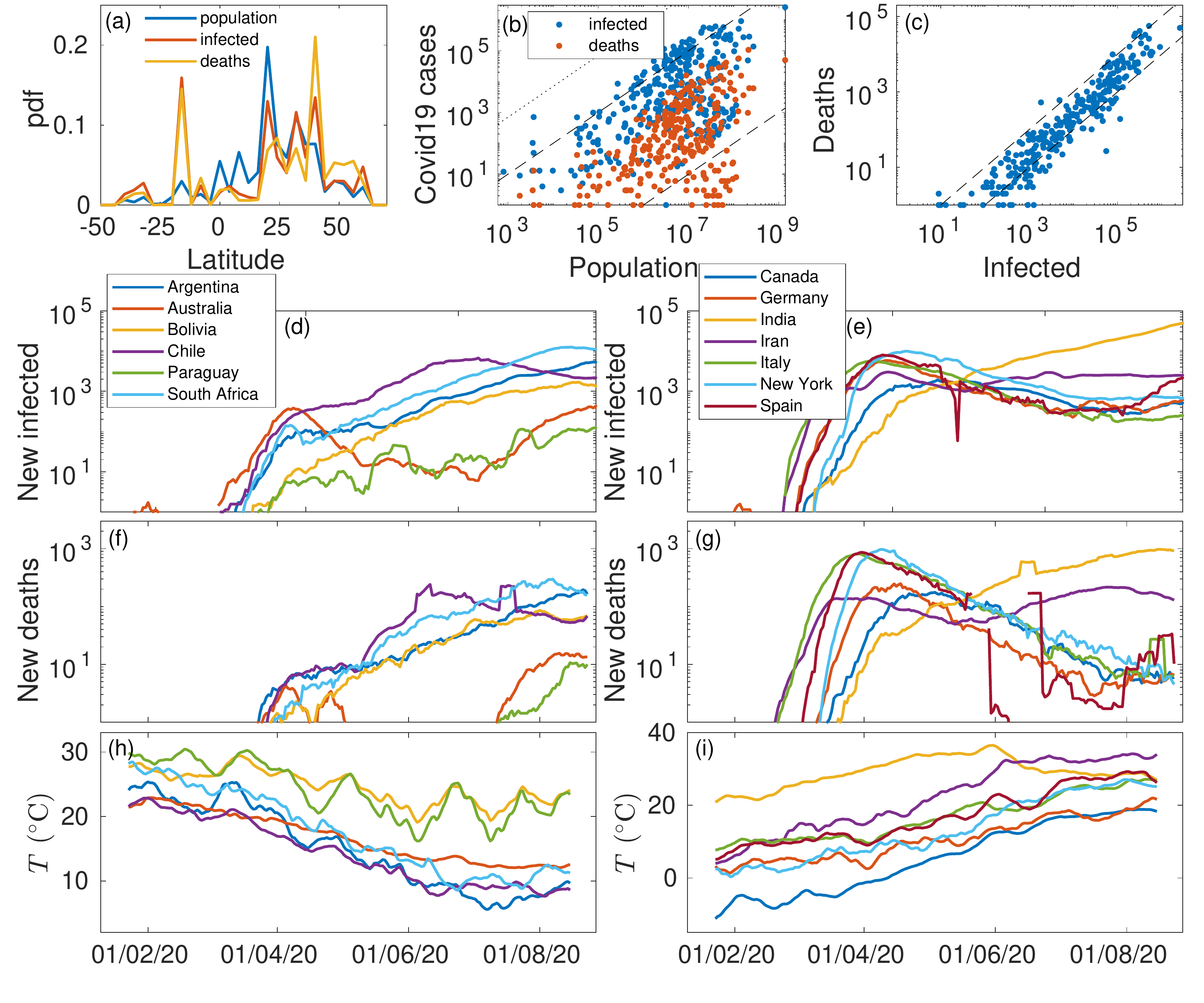}
	\caption{{\bf World population, COVID-19 cases, and temperature.} {(a)} The distribution of the world's population (blue), total COVID-19 infected cases (red), and total COVID-19 death cases (orange) by latitude. {(b)} Total number of COVID-19 infected cases (blue) and death cases (red) versus population. The dashed lines are reference lines of slope 1 indicating the 1\% (upper line) and 10$^{-4}$\% (lower line) percentages, and the dotted line indicates the 70\% herd immunity value. Note the mismatch between the population and COVID-19 cases, suggesting that additional factors other than population size are affecting the disease spread. {(c)} COVID-19 death cases versus COVID-19 infected cases. The dashed lines are reference lines of slope 1 indicating the 10\% (upper line) and 1\% (lower line) percentages. Note the large spread between the two, covering more than 1 order of magnitude. {(d-e)} One-week moving average of new daily COVID-19 infected cases as a function of time (22/1/2020 to 22/8/2020) in countries from (d) the southern hemisphere (SH) and (e) the northern hemisphere (NH). {(f-g)} Same as (d-e) for the new death cases. 
{(h-i)} The corresponding one-week moving average of surface air daily mean temperatures of the countries plotted in panels d and e. It is apparent that in the NH (panels e,g, and i), the number of new COVID-19 cases usually decreases as the temperature increases, while the opposite occurs in the SH (panels d,f and h). Yet, there are counterexamples in both hemispheres, (e.g., Australia in the SH and India in the NH), indicating a non-trivial relation between the spread of COVID-19 and temperature. We show a sample of countries in which the population is higher than 1 million and the number of confirmed COVID-19 cases is larger than 10,000. The temperature data is obtained from the ERA5 reanalysis\cite{hoffmann2019era,hersbach2020era5}.}
	\label{fig:density}
\end{figure}

The patterns of human mobility and their effect on the spread of COVID-19 have been studied for the initial stage of the pandemic\cite{gross2020spatio}. In the early stage of the virus's spread in China, it seemed as if the spread could be solely explained by the mobility patterns; i.e., L\'evy flight\cite{gonzalez2008understanding} patterns. These patterns are characterized by many short movements, together with a few very long range movements according to a power law distribution. Yet, human mobility patterns alone cannot explain the virus's continued spread to other regions around the globe.

This is mainly because the spread of the virus is not evenly distributed in different countries and does not follow the distribution of population around the globe. This mismatch can be easily seen by comparing the population distribution to the distribution of the COVID-19 (infected and death) cases (Fig.~\ref{fig:density}a). This can also be seen by plotting the number of COVID-19 cases versus population (Fig.~\ref{fig:density}b)---while there is a general increase in COVID-19 cases as a function of population size, the range of (infected and death) cases spans several orders of magnitude, indicating a mismatch between the population size and COVID-19 spread. Moreover, the number of confirmed COVID-19 cases is not closely related to the number of COVID-19 deaths as can be clearly seen from spread of points between the two dashed lines in Fig.~\ref{fig:density}c, covering more than one order of magnitude. This large spread is probably due to biases associated, for example, with in the number of COVID-19 tests which vary with time and from one country to another.

One possible explanation for the mismatch between COVID-19 cases and population is related to atmospheric conditions, about which there is an intense debate\cite{caspi2020climate, oliveiros2020role, luo2020role,wang2020temperature}. It is apparent that most of the northern hemisphere (NH) countries have passed the first wave of the disease and are currently recovering from COVID-19 spread (Fig.~\ref{fig:density}e,g), as they progress through the NH summer (Fig.~\ref{fig:density}i), while southern hemisphere (SH) countries experienced a delayed burst of COVID-19 (Fig.~\ref{fig:density}d,f) with an increasing number of COVID-19 cases during the SH winter.
This suggests a possible link between temperature (and other climate variables) and the spread of COVID-19. Yet, there are counterexamples in both hemispheres (e.g., India in the NH and Australia in the SH), which complicate the study of the correlation between climate variables and the spread of COVID-19. 

It has been shown\cite{chan2011effects} that high temperatures ($>$38 $^\circ$C) and high relative humidity ($>95 \% $) disrupt SARS-CoV viability and activation. It is also known that other types of viruses, such as influenza, have higher activation levels in colder weather; the reason for this, however, is still under debate\cite{foxman2015temperature, cannell2006epidemic}. Currently, during the NH summer, it seems that the seasonal cycle does not affect COVID-19 in the way it affects influenza.

The effect of the atmospheric conditions on COVID-19 spread is still under debate. The origin of this debate is related to many biases in the COVID-19 data, which make it very difficult to compare one place to another. The biases include different healthcare capabilities in different countries, different numbers of COVID-19 tests administered in different countries and different time, partial or possibly intentionally incorrect information published by some countries, different age pyramids, and different countermeasures and human mobility restrictions. Moreover, some studies reported a high COVID-19 replication rate under colder conditions\cite{caspi2020climate}, while other studies claimed that infection rates increase only with temperature and are negatively correlated with humidity\cite{oliveiros2020role}. Similarly, a study of 429 cities in China\cite{wang2020temperature} found an increased risk of spread in a narrow temperature range and that both high and low humidity rates are associated with higher reproduction rates\cite{luo2020role}.

Most of the studies on the effect of climate conditions on the spread of COVID-19 have concentrated on temperature and humidity and are mainly on limited regional scale. However, UV radiation has received much less attention. First, we note that the virus causing COVID-19 can live for almost three days on surfaces\cite{van2020aerosol} and also that other corona-viruses are highly sensitive to UV light\cite{iddrisu2020effects, darnell2004inactivation}. Artificial disinfection by UV light takes about 15 minutes, and UV radiation is commonly used as a germicidal disinfectant, both directly and indirectly. For example, experiments investigating the effectiveness of non-direct artificial UV lights that were installed in hospital rooms reported a reduction in tuberculosis of almost 80\%\cite{mphaphlele2015institutional}. While most artificial UV disinfection lights range from 250–305 nm\cite{walker2007effect,lytle2005predicted,battigelli1993inactivation,garland1990geographic,zak2002role} and 250 nm wavelengths barely reach ground level, the effectiveness of virus disinfection by longer wavelength UV lights (that do reach ground level and penetrate the atmosphere) is proven by SODIS (SOlar water DISinfection)\cite{heaselgrave2006solar, harding2012using, mcguigan2012solar,wegelin1994solar}. This method of water disinfection is effectively used to disinfect water against the rhinovirus (common cold), polio virus, and norovirus. It is used in the developing world daily, for water purification, in more than 2 M houses. In contrast to the 15-minute disinfection period, it disinfects water by exposing it to UV from the sun for over 12 hours. Indeed, a recent modeling study suggested reduced COVID-19 infection during the summer due to the relatively high sunlight UV radiation at ground level\cite{sagripanti2020estimated}. 

Following the above summary, the goal of this study is to quantify the effects and significance of climate variables (temperature, specific and relative humidity, and UV radiation) on the spread of COVID-19, using surrogate data tests. The proposed tests are based on a random shuffling of climate records from different locations on the globe and a comparison of the shuffled data to the original data (e.g., relative number of COVID-19 death cases). For more details, see the Materials and Methods section below. We applied the tests to both global data and data from individual regions. Our results indicate a significantly high number of COVID-19 cases (above the 95\% confidence level of the shuffled surrogate control) when the temperature is about 10$^\circ$C, the relative humidity is $\sim$60\% the specific humidity is $\sim$5 g/kg, and the UV radiation is $\sim$ 80 kJ/m$^2$.

\section*{Results}

\begin{figure}[htb!]
	\centering
	\includegraphics[width=\textwidth]{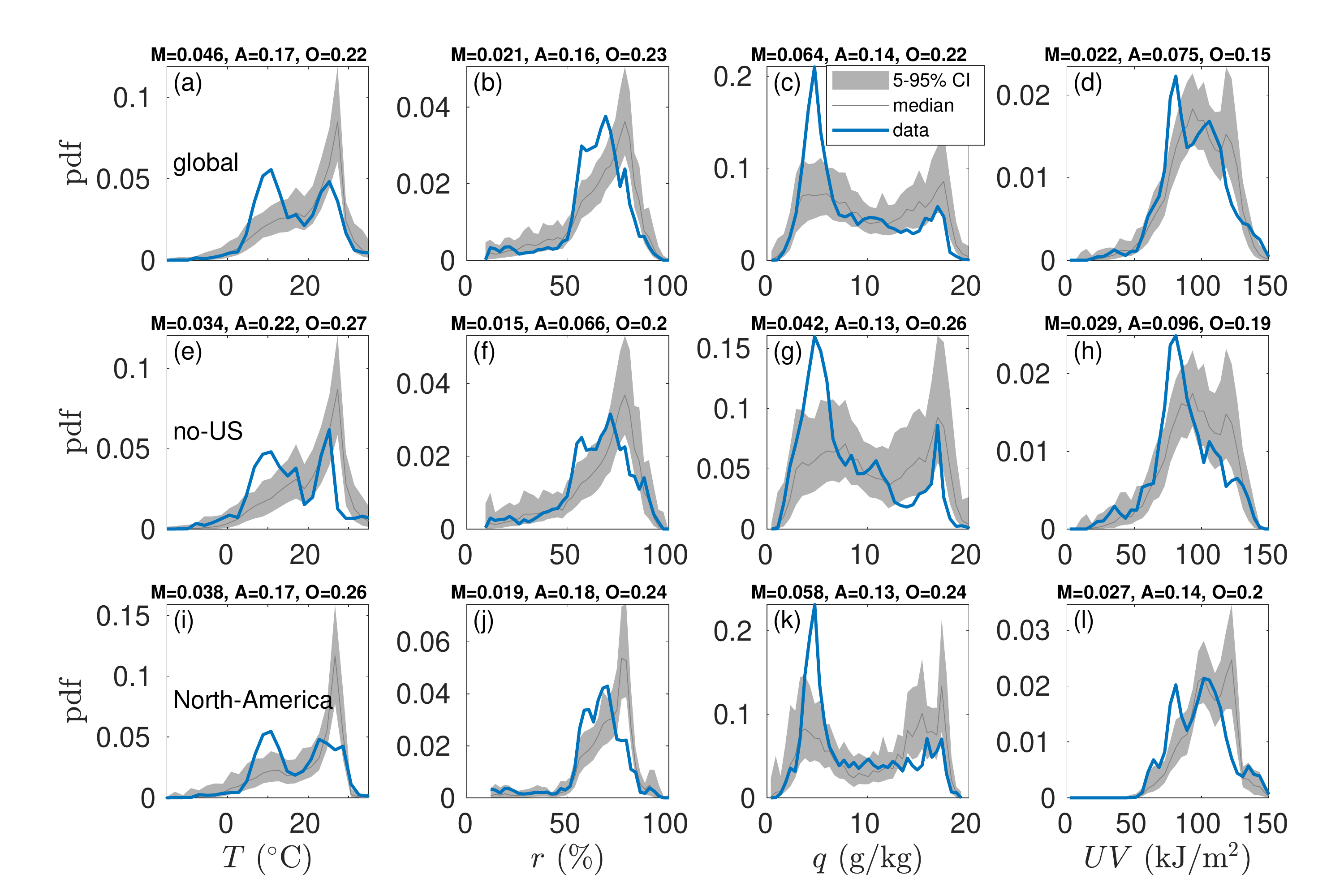}
	\caption{ {\bf The real and surrogate data pdfs of COVID-19 relative deaths cases as a function of temperature, $T$, relative humidity, $r$, specific humidity, $q$, and UV radiation.} The different rows depict the results of different regions as follows: (a-d) global, (e-h) global excluding US, and (i-l) North-America. Columns (from left to right) present the pdf versus temperature, relative humidity, specific humidity, and UV radiation. The blue curves depict the number of daily confirmed new COVID-19 cases recorded over the period of 23/1/20–15/8/20 as a function of seven-day lagged daily mean climate variables. The figure also depicts the corresponding distributions of the shuffled (location) surrogate data where the median (solid black line) and the 5-95\% confidence interval (shaded gray area) are plotted. Importantly, the peaks for temperature around 10$^\circ$C and specific humidity around 5 g/kg are significant, falling well above the 95\% surrogate level, suggesting that the COVID-19 virus has a tendency to be more effective at these temperature and specific humidity values. The title in each panel indicates the values of the separation measures between the original data pdf and the surrogate data pdf: M: maximum probability difference between the pdfs; A: the area between the original pdf (blue curve) and the confidence interval of the surrogate data pdfs (gray area); and O: the 1 minus overlap between the original pdf and the mean pdf of the surrogate data. The larger the separation value (of all measures), the greater the separation.}
	\label{fig:histograms}
\end{figure}

We summarize the main results in Fig.~\ref{fig:histograms} where we plot the probability density function (pdf) of the relative number of COVID-19 death cases (i.e., number of death cases per thousand) as a function of temperature (first column), relative humidity (second column), specific humidity (third column), and UV radiation (forth column). The different geographic regions include the entire globe (first row), the globe excluding the US (second row), and North-America (third row). We used the daily mean climate data of a 14-day backward mean. The results are consistent over the different regions and suggest a preference for spread in a relatively narrow range of specific humidity. The maximum of the probability density is well above the 95\% confidence level for a relatively low temperature (around 10$^\circ$C) and low humidity (around 60\% relative humidity and 5 g/kg specific humidity). In some cases, it is below the 5\% confidence level for higher temperature and higher (relative and specific) humidity. One possible explanation for the cold and dry weather preference may be the virus's poor viability in higher temperatures and humidity. As shown in ref. \cite{chan2011effects}, SARS-CoV's survival and activation levels in high temperatures and high humidity are also poor. Another explanation for the cold/winter weather preference may be the relatively lower levels of UV radiation from the sun during this season. 

\begin{figure}[htb!]
	\centering        \includegraphics[width=\textwidth]{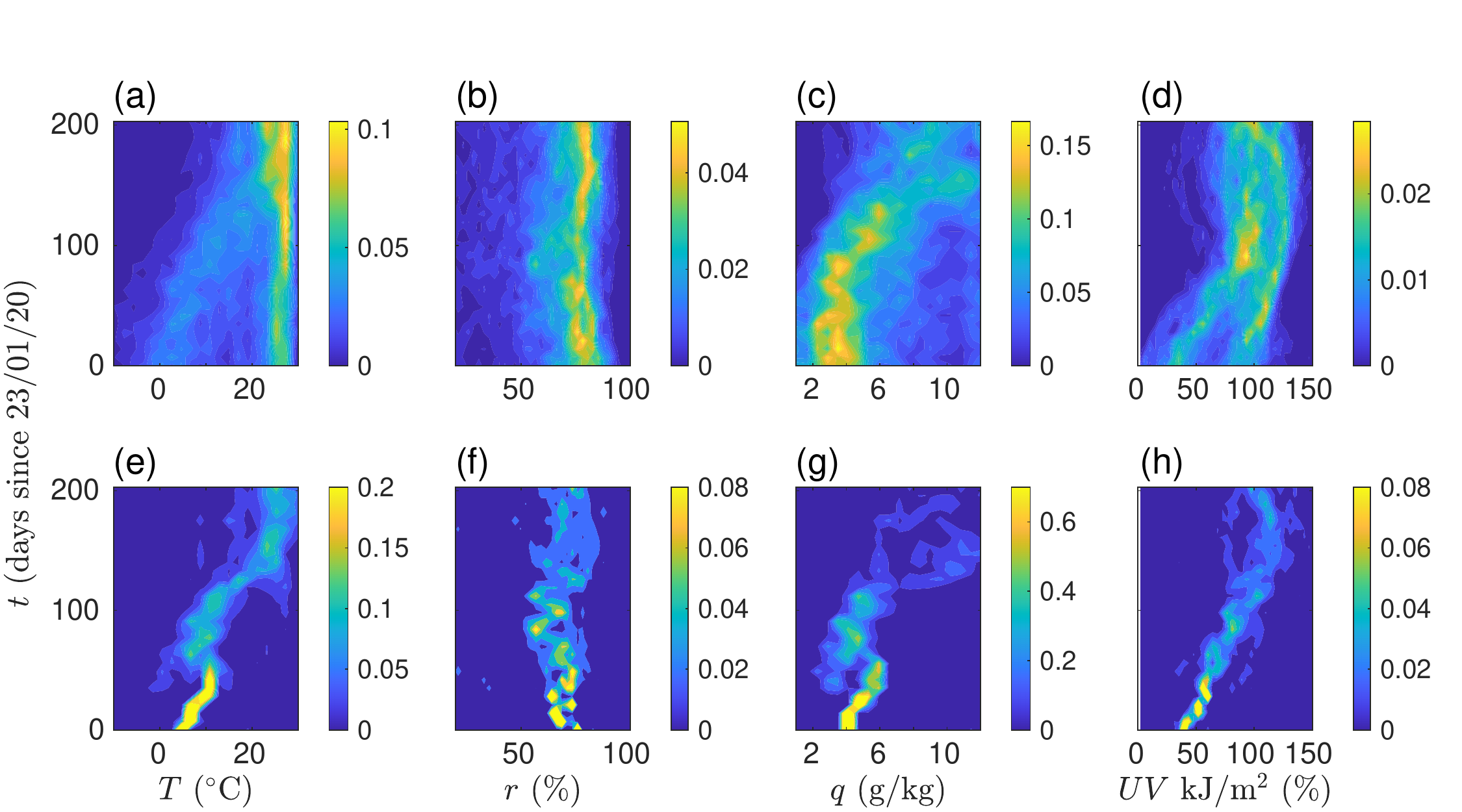}
        \caption{{\bf Evolving pdfs of climate variables and COVID-19 death cases.} Upper panels: The pdf (colors) of temperature, relative humidity, specific humidity, and UV radiation as a function of time (1/23/20–15/8/20), based on the reported locations of COVID-19 death cases around the world. The number of cases is not taken into account here so that the results reflect the changes in temporal atmospheric variables from mainly NH winter conditions (as most of the locations of the COVID-19 cases are from the NH) to mainly NH summer conditions. This seasonal trend is seen in (a), (c), and (d) but is most clearly reflected in the UV radiation (panel d), which increases toward the NH summer solstice and decrease afterward. The relative humidity does not exhibit seasonal trend. Lower panels: Pdfs (colors) of COVID-19 relative death cases versus time and temperature, relative humidity, specific humidity, and UV radiation. Here we use a daily mean and 14-day backward mean climate variables. }
	\label{fig:weeklyMeans}
\end{figure}

The analysis of UV radiation is shown in the fourth column of Fig.~\ref{fig:histograms} and indicates that the number of COVID-19 relative death cases is above the 95\% confidence level when the UV radiation is $\sim$80 kJ/m$^2$. Since the virus lives on steel and plastic surfaces for several days\cite{bin2016environmental, van2020aerosol, de2016sars}, it is possible that sites with lower UV radiation levels will suffer from a longer survival time of the virus on surfaces, leading to higher infection rates. Also, vitamin D, which is needed for the activation of the lungs' immune system, requires UV light for its formation; thus, lower exposures to UV radiation\cite{olds2010elucidating} may reduce its production. Our results support the modeling study of ref. \cite{sagripanti2020estimated}.

The saturated water vapor pressure can be determined by using the Clausius–Clapeyron relation\cite{Marshall-Plumb-2008:atmosphere}. This relation yields an exponential relation between the air temperature and the saturated water vapor pressure. The specific humidity is a measure for the amount of water vapor in the air (e.g., in grams of water vapor per kg of air) while the relative humidity is the ratio between the measured specific humidity and the maximum specific humidity. It turns out that the relative humidity exhibits weak (if at all) seasonality such that the specific humidity is expected to have seasonal cycle with higher value during the hot season. 

Fig.~\ref{fig:weeklyMeans} shows the evolution with time of the pdfs of the temperature, relative and specific humidity, and UV radiation. The daily mean values with the 14-day backward mean are used. The preference of COVID-19 for a specific temperature and humidity can be also seen in Fig.~\ref{fig:weeklyMeans}. The upper panels present the (weekly) pdfs of temperature, humidity, and UV radiation in the locations of the reported COVID-19 cases, where the relative number of death cases is not taken into account. The seasonal trend toward warmer, more humid, and higher UV radiation levels is clearly seen. This is expected as the majority of the COVID-19 cases were reported in the NH such that the seasonal trend reflects the NH winter to summer trend. In comparison, the lower panels of Fig.~\ref{fig:weeklyMeans} present the weekly pdfs of COVID-19 relative death cases as a function of temperature, relative and specific humidity, and UV radiation. In contrast to the top panels, the COVID-19 pdfs are more skewed with a single narrow maximum that follows, to some extent, the seasonal trends shown in the upper panels. We note that the relative humidity does not exhibit a clear seasonal trend. The pdf of COVID-19 relative death cases as a function of temperature peaks around 10$^\circ$C at the first part of the pandemic and then switches to $\sim$25$^\circ$C toward the NH summer, mainly as a result of the late COVID-19 burst in India and Brazil. The situation is somehow similar for the specific humidity which peaks around 5 g/kg at the first part of the pandemic. The transition from the first part of the pandemic to the second part can also be seen by comparing the pdfs of the first three months of the COVID-19 pandemic (23/1/2020 to 22/4/2020) to the last three months (16/5/2020 to 15/8/2020) where the preferred temperature of 10$^\circ$C and the preferred specific humidity of 5 g/kg is clear in the first period but not in the second period (Figs. \ref{fig:Feb_March}, \ref{fig:May_June}).

\begin{figure}[htb!]
	\centering
\includegraphics[width=\textwidth]{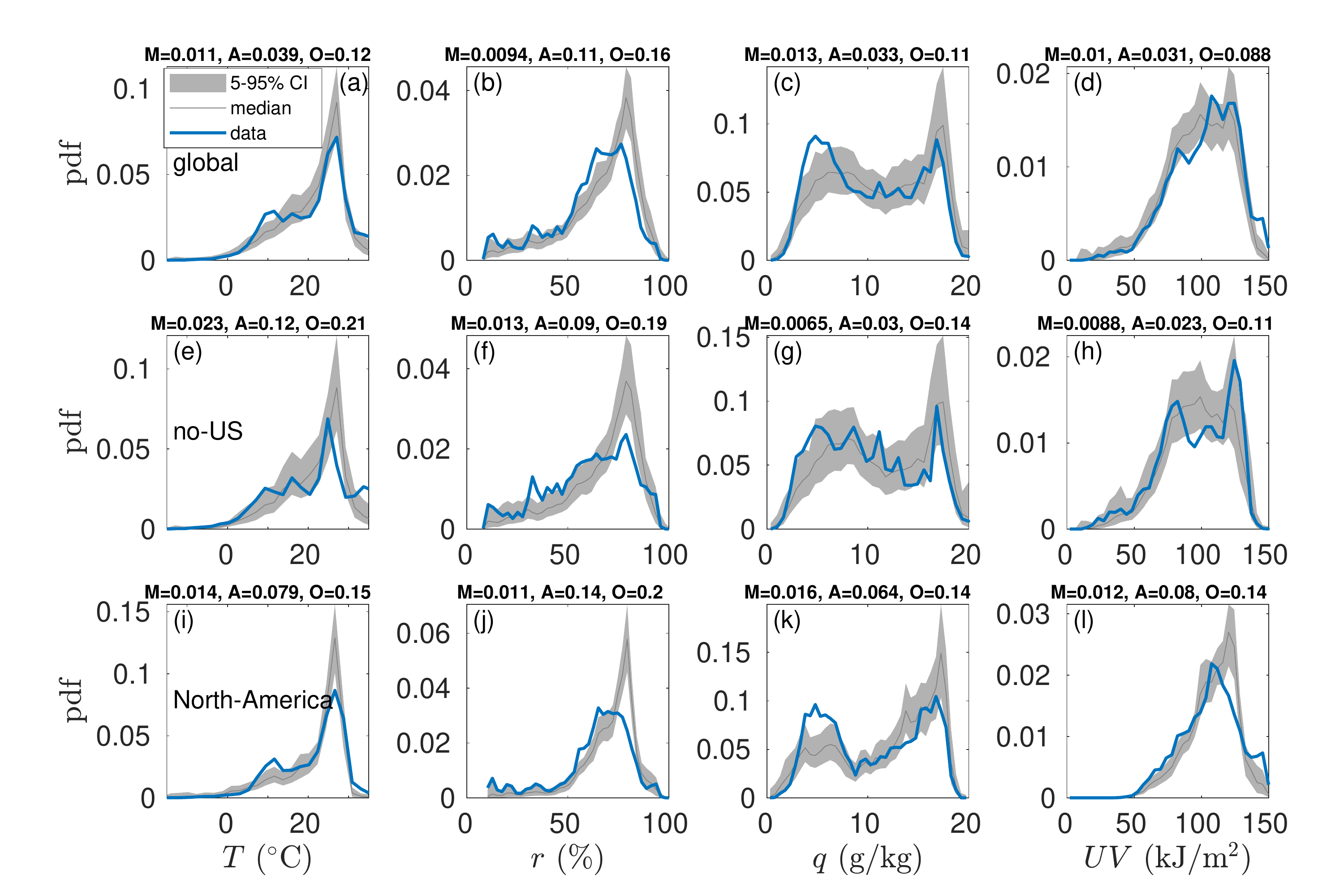}
	\caption{{\bf Results of COVID-19 infected cases.} Same as Fig.~\ref{fig:histograms} for the relative number of confirmed COVID-19 cases. Here we consider the daily mean values with a 7-day backward mean. The results show similar preferences as those of Fig.~\ref{fig:histograms}.  }
	\label{fig:confirmed}
\end{figure}

To strengthen the results reported above of enhanced COVID-19 spread for a preferred temperature, relative humidity, specific humidity and UV radiation, we also analyzed the relative number of COVID-19 confirmed cases. The results are shown in Fig. \ref{fig:confirmed} and exhibit similar results as for the relative number of COVID-19 death cases seen in Fig. \ref{fig:histograms}. Yet, the confirmed cases results are less significant, probably due to biases associated with the number of tests which become more available with time.

The results described above are based on the relative number of COVID-19 deaths and confirmed cases, i.e., the number of COVID-19 cases divided by the size of population. The normalization by population aims in taking into account, very roughly, the population density. Still, it is interesting also to check the results without the normalization by the population size, as clearly in most countries the situation is far from the herd community ratio (see Fig. \ref{fig:density}a). The results are shown in Fig. \ref{fig:no_pop_norm} and are similar to the results when the normalization by population is performed (Fig. \ref{fig:histograms}). This strengthen our conclusion regarding the preference of the COVID-19 virus to develop under cold and dry conditions. We also analyzed the number of confirmed COVID-19 cases without normalization by population and here there is no significant preference to specific value of temperature and humidity (Fig. \ref{fig:confirmed_no_pop_norm}). This is probably due to the biases associated with the number of COVID-19 tests versus time.

In addition to the surrogate method presented here, we performed two other surrogate data tests to support the effect of climate variables on the spread of COVID-19; for more details, see the Materials and Methods section. In the first surrogate test, we randomly chose the longitudes (over land) but kept the original latitudes of the COVID-19 cases. This aims to find a possible preference of COVID-19 for specific climate conditions, while keeping the seasonality (as the original latitude is maintained). In the second, additional, surrogate approach, both the longitudes and latitudes were chosen randomly where the random locations are evenly distributed over land. This aims to test the spread of COVID-19 with respect to global (continental) mean climate conditions. The results of these methods are shown in the SI (Figs. \ref{fig:shuf2},\ref{fig:shuf3}) and indicate that COVID-19 spread is affected by temperature, relative and specific humidity, and UV radiation, supporting the results reported above.

\section*{Discussion}

The fast spread of COVID-19 has resulted in a global turbulence of fear, uncertainty, and social distress. In some countries, severe quarantine restrictions were imposed, changing people's lives entirely. Then, the restrictions were slowly removed. Yet, even the most infected places in the world are far from reaching the herd immunity percentage\cite{altmann2020policy, kwok2020herd}; see the dotted line of Fig. \ref{fig:density}b. While it is clear that the most effective strategy to fight the spread of COVID-19 is quarantine restrictions, our results indicate that high temperature, high specific humidity, and high UV radiation can slowdown the spread of the disease. Our results might explain the delayed burst of COVID-19 in some warm climate countries and SH countries which entered the winter when the NH entered the summer. We conjecture that the situation could have been much worse if the climate in countries such as India and Brazil, which experienced late bursts of COVID-19, was colder and drier. Moreover, it is plausible that the climate, which is warmer humid during the NH summer, helped to cope with the pandemic in many places. If our conjectures are correct, the spread of COVID-19 will be enhanced toward the next NH winter in the absence of effective medication or vaccine.

The implications of our results demand attention. In the current period in the month of September it seems as though most NH countries already exit the first wave of the pandemic while some countries experience now the second wave of the pandemic; the most infected country, the USA, is still in the midst of its fight. In contrast, SH countries (excluding, e.g., New Zealand) are manifesting a growth in their infection curve (see Fig. \ref{fig:density}). We suggest that since the SH countries are now in winter, COVID-19 spreads more easily due to the lower temperatures and lower UV radiation. 

\section*{Data and Methods}

\label{sec:data}

We extracted climate variables from the ERA5 reanalysis database\cite{hoffmann2019era,hersbach2020era5}; ERA5 is a high spatial (1/4 of a degree) and temporal (hourly) resolution database that includes many multi-level climate variables. We focus on surface level data of 2~m temperature, downward UV radiation (in the range of 200–440 nm) at the surface, and 1000 hPa relative and specific humidity. We tested whether these climate variables can be associated with the spread of COVID-19. We used the hourly data to extract the daily mean and daily maximum values of the different climate variables.

The COVID-19 data is obtained from the Johns Hopkins COVID-19 GitHub repository\cite{CSSE}, and the demographic data is obtained from\cite{NASA}. The COVID-19 data includes the number of confirmed cases, the number of active cases, the number of severe cases, the number of deaths, and more. Here, we focused on the daily deaths and the number daily confirmed new cases. We used the daily deaths, along with the daily confirmed cases, since presumably there are infected people that are not tested, but this happens less often for deaths. Furthermore, the number of tests differ between developed and developing countries, affecting the reported number of infections, while the number of deaths is not subject to this bias. The data is provided for the entire country (e.g., Germany and Italy), for different provinces and regions within a country (e.g., UK, Canada, China, and Australia), and for cities within a country (such as the US). We used the regional COVID-19 data when possible.

We mainly analyzed the ``normalized'' number of death cases by considering the number of cases per 1000 inhabitants. We concentrated on countries/states/provinces whose population is larger than half a million. The population normalization was performed in order to filter out the population size effect, and we found results that are similar to the results without population normalization; see Figs. \ref{fig:histograms}, \ref{fig:no_pop_norm}. In addition to the analysis of daily new COVID-19 death cases, we also considered the number of daily confirmed COVID-19 cases.

The different climate variables were interpolated to the reported locations of COVID-19 cases. Then, for each location and date, we calculated the past $d$-day mean (either of the daily mean or daily maximum) and the $d$-day lag values of the climate variables. The rationale behind the $d$-day mean operation is that new COVID-19 death cases may occur after a variable number of days, somewhere between a few days (from the infection time) to more than one month--the mean operation crudely reflects this temporal spread. The $d$-day lag operation aims to examine the other extreme, unrealistic alternative in which new COVID-19 death cases occur after a fixed number of days after the infection. The daily-mean and daily-max procedures aim to test whether the extreme values of temperature/humidity/UV affect the virus spread or the accumulated daily value (which is reflected by the mean operation). We examined different time lags and temporal mean periods and found a typical span time of 14 days (not shown). The typical span time for the number of confirmed cases is seven days. Generally speaking, the daily mean $d$-days mean procedure yielded the best results.

We developed surrogate data tests (inspired by ref. \cite{theiler1991testing}) to study whether COVID-19 spread favors a certain range of climate attributes. The common practice is to assume a NULL hypothesis and to design statistical tests that will either falsify or confirm this hypothesis. In our case, the NULL hypothesis is that the climate attributes are not related to COVID-19 spread. If this is indeed the case, the spread should not be affected by the climate conditions of a certain location. Thus, we shuffled the locations of the reported COVID-19 cases, keeping the time series of the cases unaffected but using the climate data of other random locations. The shuffling operation can be repeated many times. If the NULL hypothesis is valid, the resulting distribution of the shuffled data should be similar to the distribution of the original data. If it is significantly different than the original distribution, the NULL hypothesis is rejected, and the climate variable is proven (to a certain confidence level) to affect the spread of COVID-19.

We proposed and implemented three methods to generate random locations of COVID-19 cases. In the first method discussed above, the reported locations of the virus cases were shuffled, i.e., we analyzed the number of COVID-19 cases that were recorded in a particular location, with climatic records from a different, randomly selected location, from the list of the locations of reported cases; see Fig. \ref{fig:histograms}. In the second method (see Figs. \ref{fig:shuf2} of the SI), we randomly chose the longitude of the reported COVID-19 locations but kept the latitude; the random longitudes were restricted to be over land. In this way, we kept the original seasonality, yet studied the sensitivity of the results to the climate variability along the original latitude. In the third method, we randomly chose both the longitude and latitude and used the climate records of the random locations instead of the climate records of the original locations. The new random locations were chosen to be evenly distributed over land.

We constructed the probability density function (pdf) of the COVID-19 cases using the various climate variables; see Fig. \ref{fig:histograms}. Then, the pdf of the original data was compared to the pdf of the surrogate data. We generated 200 surrogate time series for each reported COVID-19 location and then estimated the 5\% and 95\% confidence level. High separation between the pdf of the real data and the pdf of the surrogate data implies a higher dependency of COVID-19 on the climatic variables. We quantified the dissimilarity between the pdfs of the real data and the surrogate data in three ways: (a) the maximum probability difference between the real data pdf and the 95\% confidence level of the surrogate data pdf, (b) 1 minus the overlapping area between the original data pdf and the mean pdf of the surrogate data, and (c) the sum of the areas that are bounded between the original data pdf and the 5\% and 95\% confidence level of the surrogate data pdfs (i.e., the area between the blue curve in Fig. \ref{fig:histograms} and the shaded gray area). We use the letters ``M,'' ``O,'' and ``A,'' respectively, to refer to these three separation measures. The larger the separation value (of all measures), the greater the separation.




\section*{Author contributions statement}

E.E.A. performed the initial analysis. Y.A. suggested the statistical approach, performed the statistical analysis, and prepared the figures. A.S. contributed the UV idea. E.E.A., Y.A., S.H., and A.S. reviewed the manuscript.

\section*{Additional information}


The authors declare no competing (financial and non-financial) interests. Correspondence and requests for materials should be addressed to Y.A. (ashkena@bgu.ac.il).


\setcounter{figure}{0}
\renewcommand{\thefigure}{S\arabic{figure}}

\section*{Supporting information}
\begin{figure}[h]
	\centering
\includegraphics[width=\textwidth]{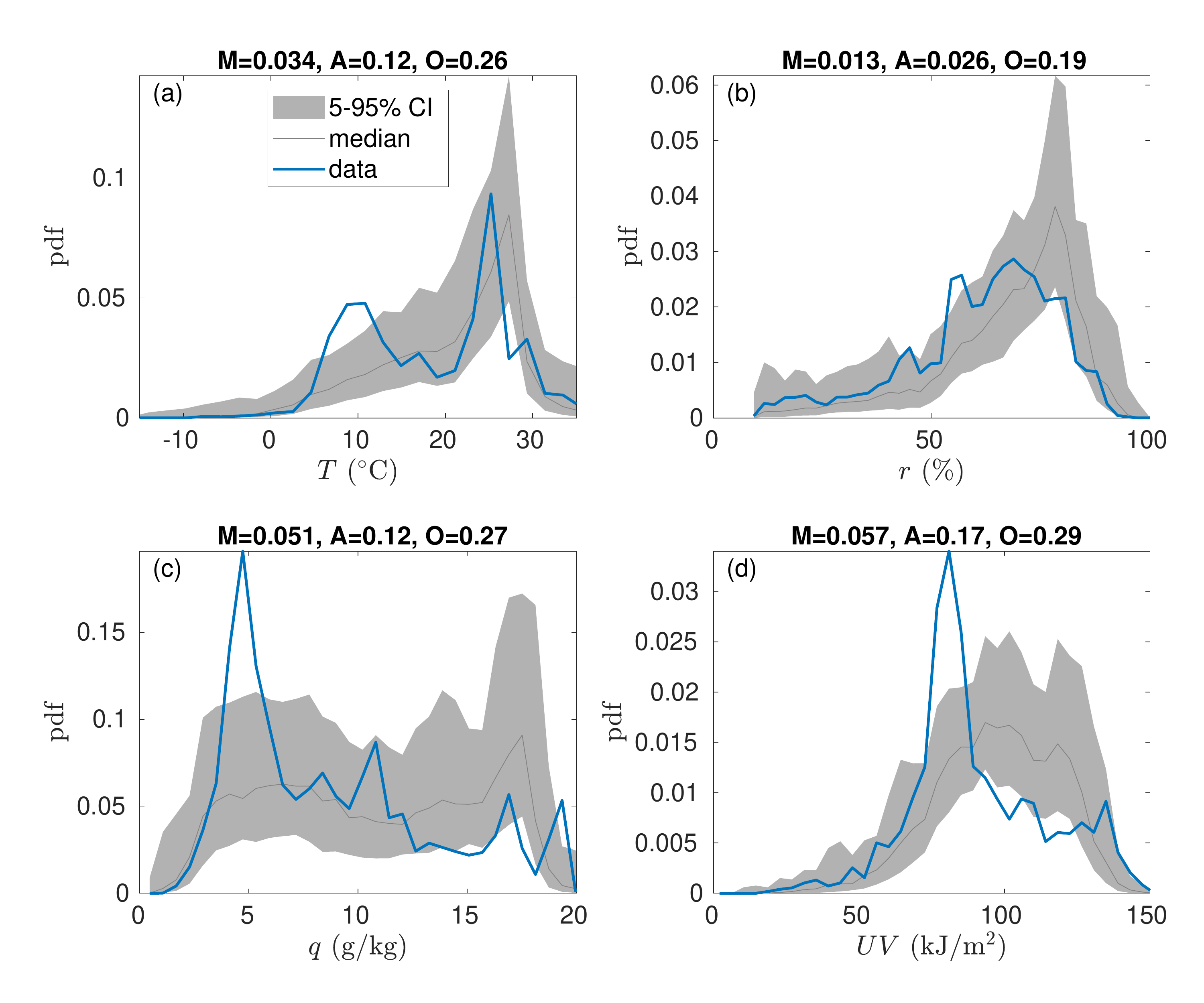}
	\caption{ {The real and surrogate data pdfs of COVID-19 death cases as a function of (a) temperature, $T$, (b) relative humidity, $r$, (c) specific humidity, $q$, and (d) $UV$ radiation.} Here we used global data from 23/1/2020 to 15/8/2020 without normalization by the population size.}
	\label{fig:no_pop_norm}
\end{figure}

\begin{figure}[h]
	\centering
\includegraphics[width=\textwidth]{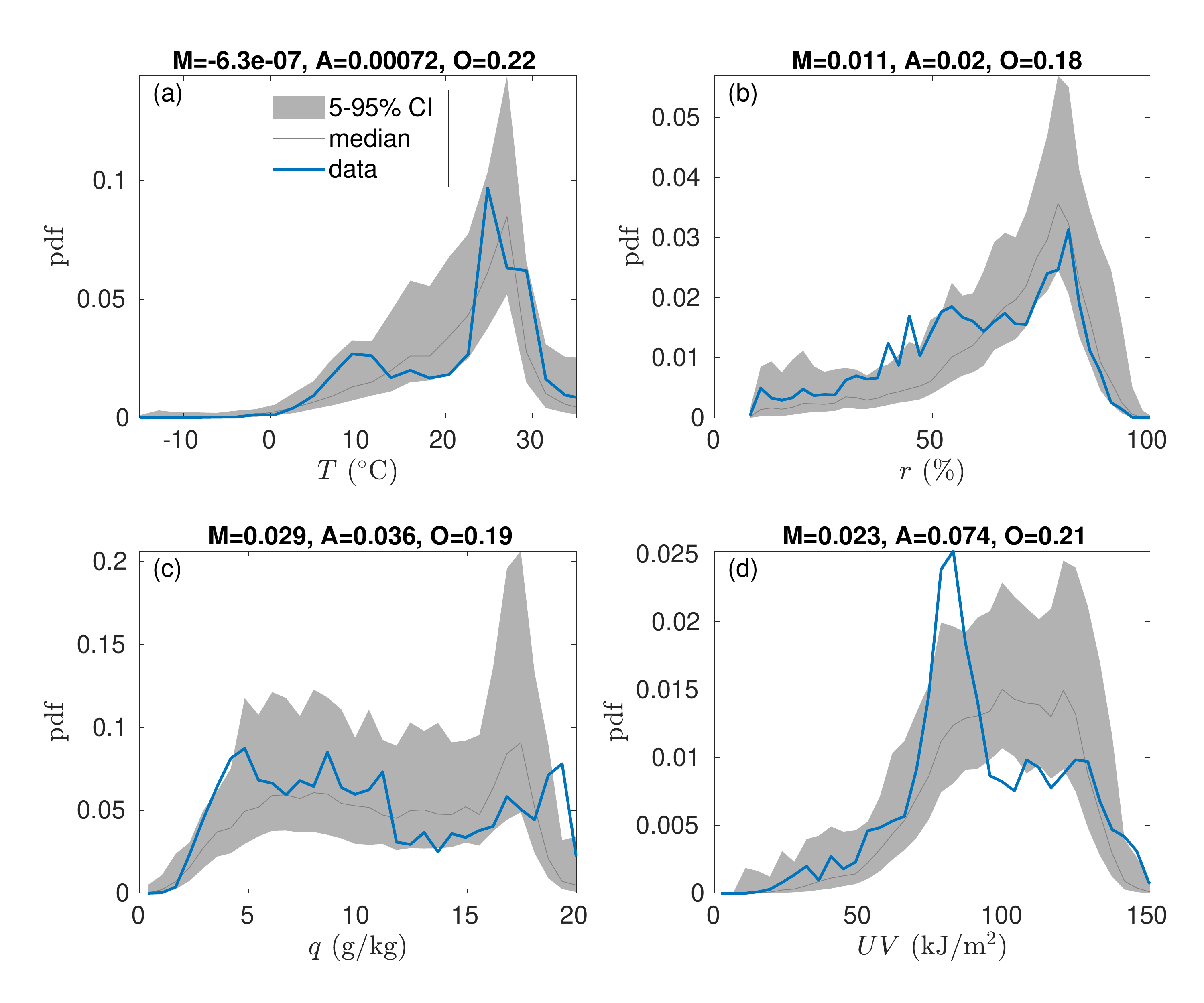}
\caption{ {The real and surrogate data pdfs of COVID-19 confirmed cases as a function of (a) temperature, $T$, (b) relative humidity, $r$, (c) specific humidity, $q$, and (d) $UV$ radiation.} Here we used global data from 23/1/2020 to 15/8/2020 and the number of confirmed (without normalization) COVID-19 cases by the population size.}
	\label{fig:confirmed_no_pop_norm}
\end{figure}

\begin{figure}[h]
	\centering
	\includegraphics[width=\textwidth]{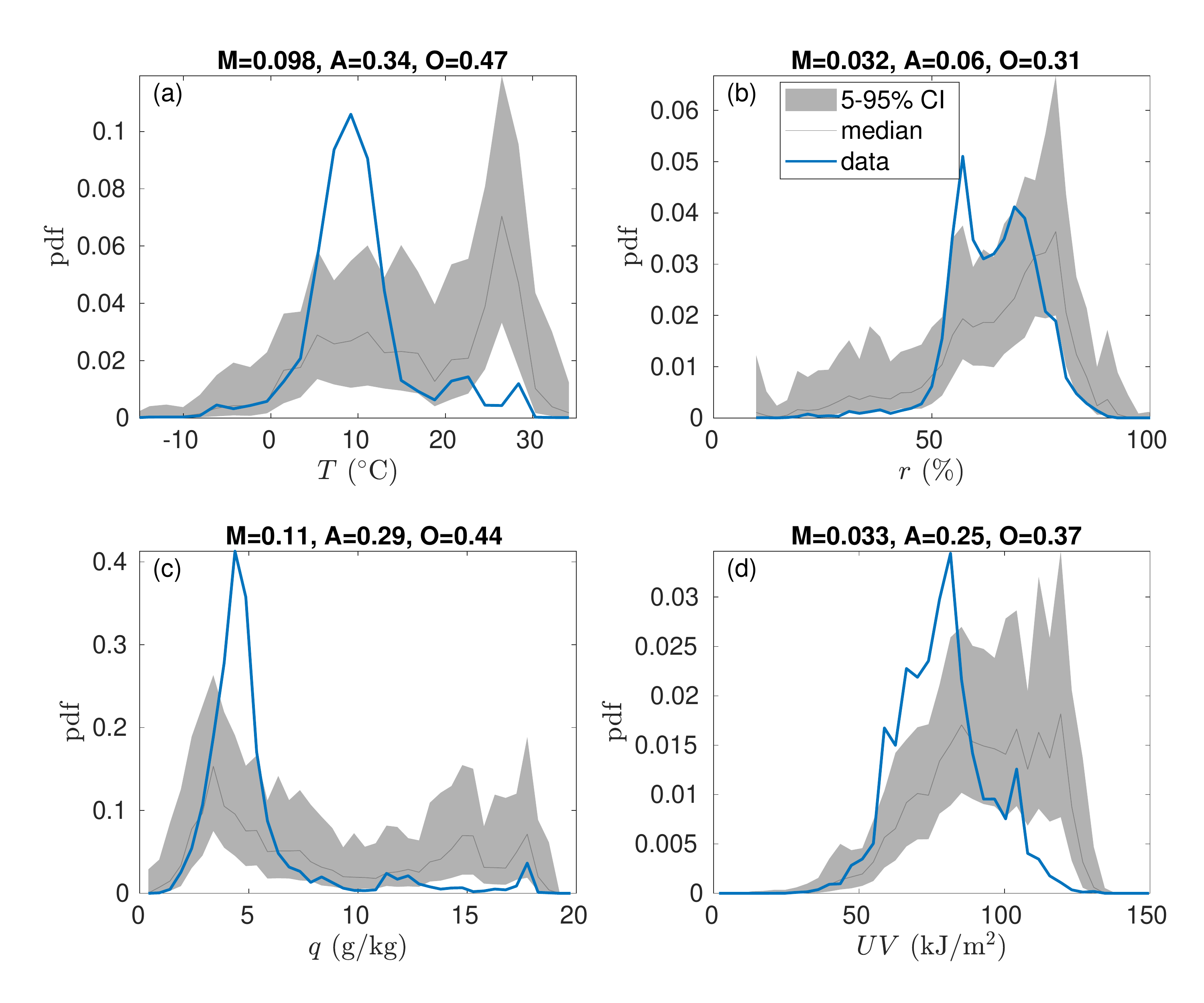}
	\caption{ Same as Fig.~\ref{fig:no_pop_norm} with population normalization and during the first three months of the COVID-19 pandemic, from 23/1/2020 to 22/4/2020.}
	\label{fig:Feb_March}
\end{figure}

\begin{figure}[h]
	\centering
\includegraphics[width=\textwidth]{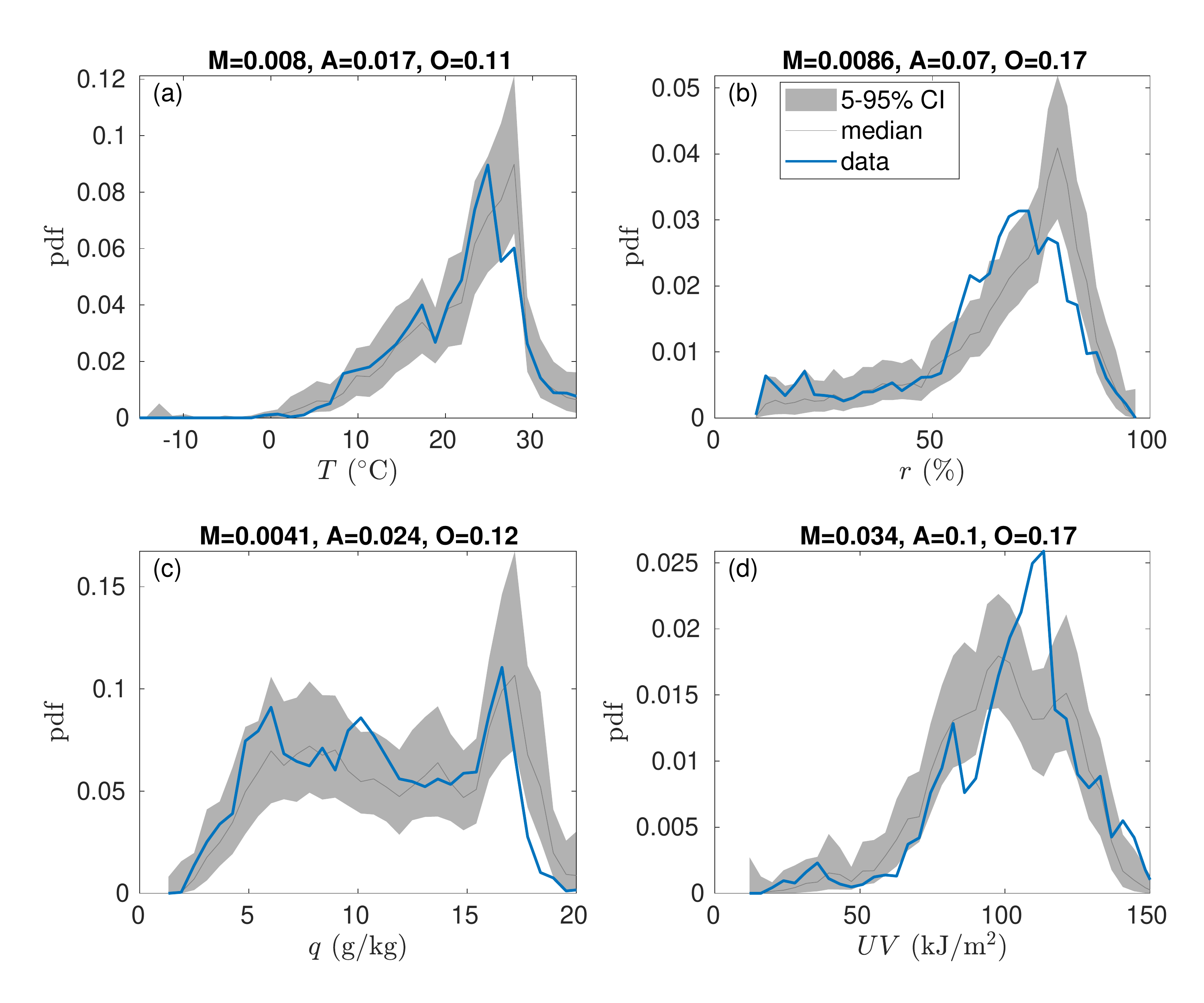}
	\caption{ Same as Fig.~\ref{fig:Feb_March} for the last three months of the COVID-19 pandemic, from 16/5/2020 to 15/8/2020.}
	\label{fig:May_June}
\end{figure}

\begin{figure}[h]
	\centering
\includegraphics[width=\textwidth]{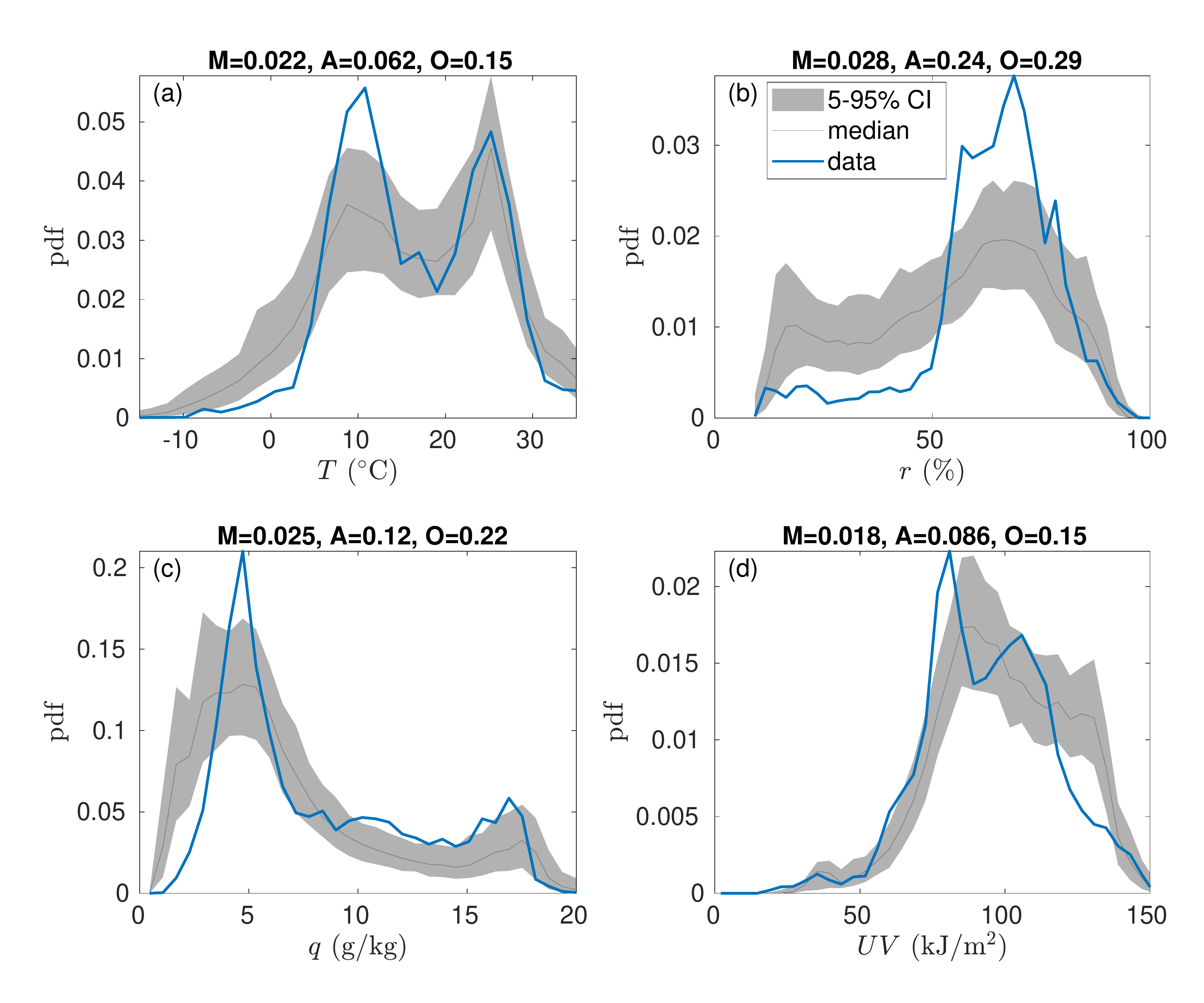}
	\caption{ Same as Fig.~\ref{fig:Feb_March} for the entire COVID-19 pandemic period, from 23/1/2020 to 15/8/2020, when using the second surrogate data test where the original latitudes are kept, but the longitudes are chosen randomly from continental areas of the specific latitudes.}
	\label{fig:shuf2}
\end{figure}

\begin{figure}[h]
	\centering
\includegraphics[width=\textwidth]{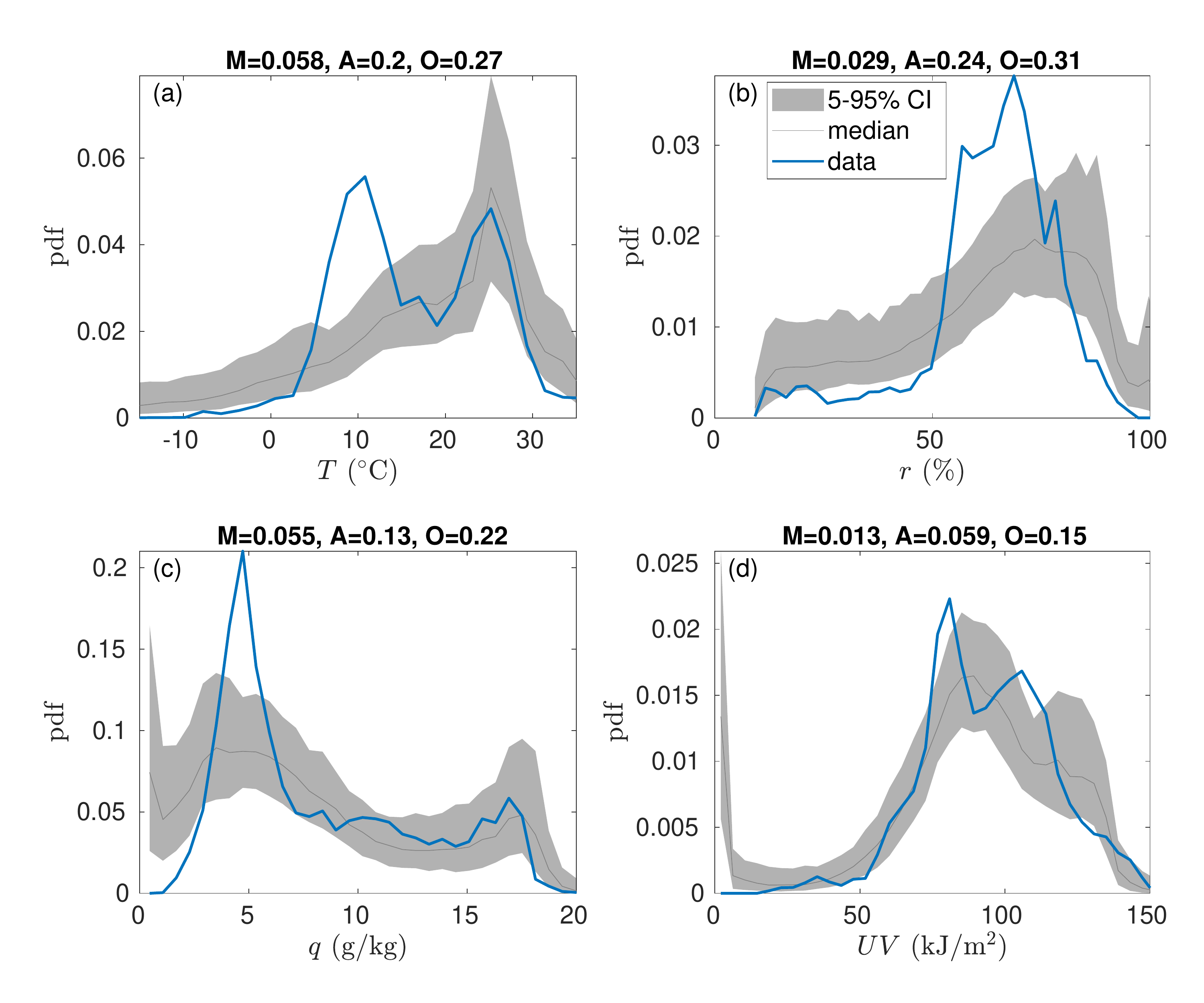}
	\caption{ Same as Fig.~\ref{fig:shuf2} when using the third surrogate data test where both the longitudes and latitudes are chosen randomly from continental areas.}
	\label{fig:shuf3}
\end{figure}

\end{document}